\documentclass[fleqn,12pt,twoside]{article}
\usepackage[headings]{espcrc1}

\readRCS
$Id: espcrc1.tex,v 1.2 2004/02/24 11:22:11 spepping Exp $
\usepackage{graphicx}
\usepackage[figuresright]{rotating}

\newcommand{\ttbs}{\char'134}
\newcommand{\AmS}{{\protect\the\textfont2
  A\kern-.1667em\lower.5ex\hbox{M}\kern-.125emS}}
\newcommand{\be}{\begin{eqnarray}}
\newcommand{\ee}{\end{eqnarray}}

\newcommand{\pr}[1]{Phys. Rev. { #1}}

\newcommand{\np}[1]{Nucl. Phys. { #1}}

\title {Status of art of reaction models for projectiles far from stability.}

\author{Angela Bonaccorso\address{Istituto Nazionale di Fisica Nucleare, Sez. di
Pisa, 
and Dipartimento di Fisica, Universit\`a di Pisa,
 Largo Pontecorvo 3, 56127 Pisa, Italy.}%
        \thanks{In collaboration with C.A. Bertulani, G. Blanchon, D.M. Brink, F. Carstoiu, A. Garc\'ia-Camacho, A.A. Ibraheem,  J. Margueron, N.Vinh Mau.} }
       
\begin{document}

\maketitle

\begin{abstract}
This talk will review the status of art of nuclear and Coulomb breakup theories and their relation to optical models of elastic scattering of exotic projectiles.
The effect of the final state interactions between the breakup particle and the core and target  nuclei will be clarified and some typical numerical calculations for  the relevant observables will be presented and compared to experimental data.
Finally new results will be shown to demonstrate the feasibility of a novel type of  experiment involving heavy projectiles far from stability on heavy targets.

\end{abstract}

\section{Introduction}

Nuclear reaction theory has experienced a great revival in the last twenty years following
the large increase in quantity as well as quality of experiments with exotic beams. Exotic nuclei
are located away from the stability valley and have large differences in the number of neutrons and protons. Their valence particle separation energies S$_n$ are smaller than the average 8 MeV expected in nuclear matter. In the extreme case of halo nuclei such as several beryllium and lithium isotopes ($^{11}$Be,$^{12}$Be,$^{14}$Be,$^{11}$Li), S$_n$ can be even less than 1 MeV. As a consequence as much as 10\% of the total reaction cross section is due to just one channel: 1n or 2n breakup.
Therefore, {\it out of necessity} (N. Orr), breakup has been  the most studied reaction for very weak beams and the one for which several new models have been developed.

In this short review I will first present the mechanisms which lead to breakup and the relative observables that are measured. For each of them a description will follow of the most recent advances in the  models used for theoretical calculations. 

\section{Cross section }

All theoretical methods used so far to describe breakup rely on a basic approximation to describe the collision with
only the three-body variables of nucleon coordinate, projectile
coordinate, and target coordinate. Thus the dynamics is controlled by the
three potentials describing nucleon-core, nucleon-target, and core-target
interactions.  In most cases the projectile-target relative motion is treated
semiclassically by using a trajectory of the center of the projectile
relative to the center of the target $\mathbf{R}%
\left(  t\right)  =\mathbf{b}_c+vt\mathbf{\hat{z}}$ with constant velocity $v$ in the $z$ direction and impact parameter
{\bf b$_c$} in the $xy$ plane.  This approximation makes our formalism applicable for incident energies above the Coulomb barrier. Along this trajectory the amplitude for a
transition from a nucleon state  $\psi_i$  bound  in the projectile, to a final  continuum state $\psi_f$, is given by \cite{adl,bb}
\begin{equation}A_{fi}={1\over i\hbar}\int_{-\infty}^{\infty}dt \langle \psi_{f} ({\bf r},t)|V({\bf { r, R}}(t))|\psi_{i}({\bf r},t)\rangle,\label{1}
\end{equation}
where $V$ is the
interaction responsible for the transition which will be specified in the following. The probabilities for different processes can be
represented in terms of the amplitude as ${dP/d
\xi}=\sum |A_{fi}|^2 \delta(\xi-\xi_f)\label{2}$ where
$\xi$ can be momentum, energy or any other variable for which a differential cross section is measured.
Direct one-particle re-arrangment reactions of the peripheral type in presence of strong core-target absorption can be described by an equation like  \cite{bb,bb4,1,bw}
\begin{equation}
{d\sigma_{-n}\over {d{{\xi_f}} }}=C^2S
\int d{\bf b_c} {d P_{b_{up}}(b_c)\over
d{{\xi_f}}}
P_{ct}(b_c), \label{cross}
\end{equation}
(see Eq. (2.3) of \cite{bb4}) and C$^2$S is the spectroscopic factor  
for the initial single particle state.
The core survival probability is defined in terms
of a S-matrix function of the core-target distance of closest approach  
$b_c$. A simple parameterisation is
 $P_{ct}(b_c)=|S_{ct}|^2=
e^{(-\ln 2 exp[(R_s-b_c)/a])}.$  
It  takes into account the peripheral nature of
the reaction and naturally excludes the possibility of large overlaps  
between projectile and target. The  
strong absorption radius R$_s\approx1.4(A_p^{1/3}+A_t^{1/3})$ fm is  
defined
as the distance of closest approach for a trajectory that is 50\%  
absorbed from the elastic channel and a=0.6 fm  is a diffuseness
parameter. The values of R$_s$ thus obtained agree within a few percent  
with those of the Kox parameterization \cite{kox}.

\section{Coulomb Breakup}

Based on the  time dependent amplitude Eq.(\ref{1}) and the classical projectile-target trajectory of relative motion  given above, in Ref.\cite{1} we considered the breakup of a halo nucleus like $^{11}$Be consisting of a
neutron bound to a $^{10}$Be core in a collision with a target nucleus.  The system of the halo nucleus and the target was described
by Jacobi coordinates $\left(  \mathbf{R,r}\right)  $ where $\mathbf{R}$ is
the position of the center of mass of the halo nucleus relative to the target
nucleus and $\mathbf {r}$ is the position of the neutron relative to the halo
core, and the coordinate $\mathbf{R}$ is assumed to move on a classical path.
The Hamiltonian of such a  system is $
H=T_{R}+T_r+V_{nc}\left(  \mathbf {r}\right)  +V_{nt}\left(  \mathbf{\beta}_{2}
\mathbf{r+R}\right)  +V_{ct}\left(  \mathbf{R-\beta}_{1}\mathbf{r}\right),
$
with  $\beta_1$ and $\beta_2$  the mass ratios of neutron and core, respectively, to that of the projectile.
 $T_{R}$ and $T_r$ are the kinetic energy operators associated with the
coordinates $\mathbf{R}$ and $\mathbf{r}$ and $V_{cn}$ is a real potential
describing the  neutron-core final state interaction. $V_{cn}$ was neglected under the hypothesis that the observables measured and calculated did not depend significantly on it. In Sec.5 we will discuss cases in which such an interaction dominates instead the measured data. The potential $V_2=V_{nt}+V_{ct}$
describes the interaction between the projectile and the target. It is a sum
of two parts depending on the relative coordinates of the neutron and the
target and of the core and the target.
  Both $V_{nt}$ and $V_{ct}$ are
represented by complex optical potentials. The imaginary part of $V_{nt}$
describes absorption of the neutron by the target to form a compound nucleus.
It gives rise to the stripping part of the halo breakup we will describe in Sec. 4. The imaginary part of
$V_{ct}$ describes reactions of the halo core with the target. The potential
$V_{ct}$ also includes the Coulomb interaction between the halo core and the
target which is responsible for Coulomb breakup.
Since the mass ratio $\beta_{1}$ is small for a halo nucleus with a heavy core ($\beta_{1}\approx0.1$ for  $^{11}$Be) the Coulomb potential was approximated by the dipole term in Ref.\cite{1}.
Then,  making   an eikonal  approximation for the neutron final state
$\psi_{f}\left(  t\right)  =\exp\left(  i\mathbf{k \cdot r}-i\varepsilon_{\mathbf{k}}%
t/\hbar\right)  \exp\left(  -\frac{1}{i\hbar}\int_{t}^{\infty}{V}%
_{2}\left(  \mathbf{r},t\right)  dt\right)  $,  the  amplitude became
\begin{equation}
A_{lm}\left(  \mathbf{k,}\mathbf{b}_c\right)  =\frac{1}{i\hbar}\int
d^{3}\mathbf{r}\int dte^{-i\mathbf{k \cdot r}+i\omega t}e^{\left(  \frac{1}{i\hbar
}\int_{t}^{\infty}{V}_{2}\left(  \mathbf{r},\mathbf{R}(t)\right)  dt\right)  }
{V}_{2}\left(  \mathbf{r},\mathbf{R}(t)\right)  \phi_{lm}\left(  \mathbf{r}\right)
\label{amp1}%
\end{equation}
where $\omega=\left(  \varepsilon_{\mathbf{k}}-\varepsilon_{i}\right)  /\hbar
$ and  $\phi_{lm}$ is the radial part of the neutron initial wave function.  $\bf k$ is the neutron-core relative momentum vector in the final state. The corresponding energy is $ \varepsilon_{\mathbf{k}}$, while it is $\varepsilon_{i}$ in the initial state. Eq.~(\ref{amp1}) is appropriate to calculate the coincidence cross section 
$A_p\to (A_p-1)+n$. 

The components $V_{nt}$ and $V_{ct}$ of ${V}_{2}$ were treated differently because of the long range of the
Coulomb interaction.  The neutron-target interaction is strong and
has a short range. We assumed that the interaction time $\tau$ for this part of the interaction is very short in the sense that
$\omega\tau$ is small compared with unity. On the other hand the the long range Coulomb interaction
between the halo core and the target is weaker and changes more slowly. The way to treat it is discussed in the next subsection.

\subsection{Coulomb phase}
Introducing the notation $V_c=Z_cZ_te^2$, $V_v=Z_vZ_te^2$ and $V_C=(Z_v+Z_c)Z_te^2$, with $Z_v=0,1$ for a neutron and  for a proton respectively,  the Coulomb potential can be written as
\begin{equation}
V({\bf r},{\bf R})= \frac{V_c}{|{\bf R}-\beta_1 {\bf r}|}+\frac{V_v}{|{\bf R}+\beta_2 {\bf r}|}-\frac{V_C}{R}.
\end{equation} Here  $\beta_1$ can be also the mass ratios of the proton  to that of the projectile.  
In Ref.\cite{nois} we have  shown that the Coulomb phase
$
 \chi_{eff}({\bf b}_c,{\bf r},k)= \int dt e^{i\omega t} V_{C}({\bf r},t)/\hbar.
$
 which is solvable in the dipole approximation, can similarly be calculated with
  the whole multipole expansion if a screening term is added and subtracted to the potential, suct that  it can be written as
$
V({\bf r},{\bf R})=V_{sh}({\bf r},{\bf R})  + V_{lo}({\bf r},{\bf R})=
V_{C} [(\frac{e^{- \gamma |\vec{R}-\beta_1 \vec{r}|}}{|\vec{R}-\beta_1 \vec{r}|} - \frac{e^{- \gamma R}}{R})+
(- \frac{1-e^{- \gamma |\vec{R}-\beta_1 \vec{r}|}}{|\vec{R}-\beta_1 \vec{r}|}+  \frac{1-e^{- \gamma R}}{R})].
$
The term $V_{sh}$ contains the singularity at R=0 but decays quickly with the impact parameter. On the other hand, $V_{lo}$, well-behaved in the origin, accounts for the long-range character of the Coulomb potential. When inserted in the integral for $\chi_{eff}$, these two terms can be treated in different ways if the parameter $\gamma$ is big enough. In this case, as done in \cite{1} with the nuclear potential, $V_{sh}$ can be considered in the sudden approximation, yielding a phase
$
\chi_{sudd}({\bf b}_c,{\bf r})=\int dt  V_{sh}({\bf r},t)/\hbar,
$
 whereas $V_{lo}$ needs to keep the whole time evolution description, but, being weak, it can be approximated to first order
$
\chi_{pert}({\bf b}_c,{\bf r},k)=\int dt e^{i\omega t} V_{lo}({\bf r},t)/\hbar.
$
 Therefore the Coulomb phase becomes a sum of two terms
$
 \chi_{eff}({\bf b}_c,{\bf r},k)= \chi_{sudd}({\bf b}_c,{\bf r})+ \chi_{pert}({\bf b}_c,{\bf r},k),
$
 both of them depending upon the screening parameter $\gamma$.  In order for this approximation to be valid, the screening term $\gamma$ needs to be sufficiently large as to ensure that the range of $V_{sh}$ remains short enough, and that $V_{lo}$ does not become too large. This is next achieved by taking just $\gamma=\infty$, in which case
$\chi_{sudd}=0$, and
$\chi_{eff}=\chi_{pert}=\frac{2 V_C}{\hbar v}\left(e^{i\beta_1 \omega z /v}K_0(\omega b_c/v) -K_0(\omega R_\perp/v)\right). $

 \begin{figure}[htb]
\begin{minipage}[t]{80mm}
\scalebox{0.27}{
               \includegraphics{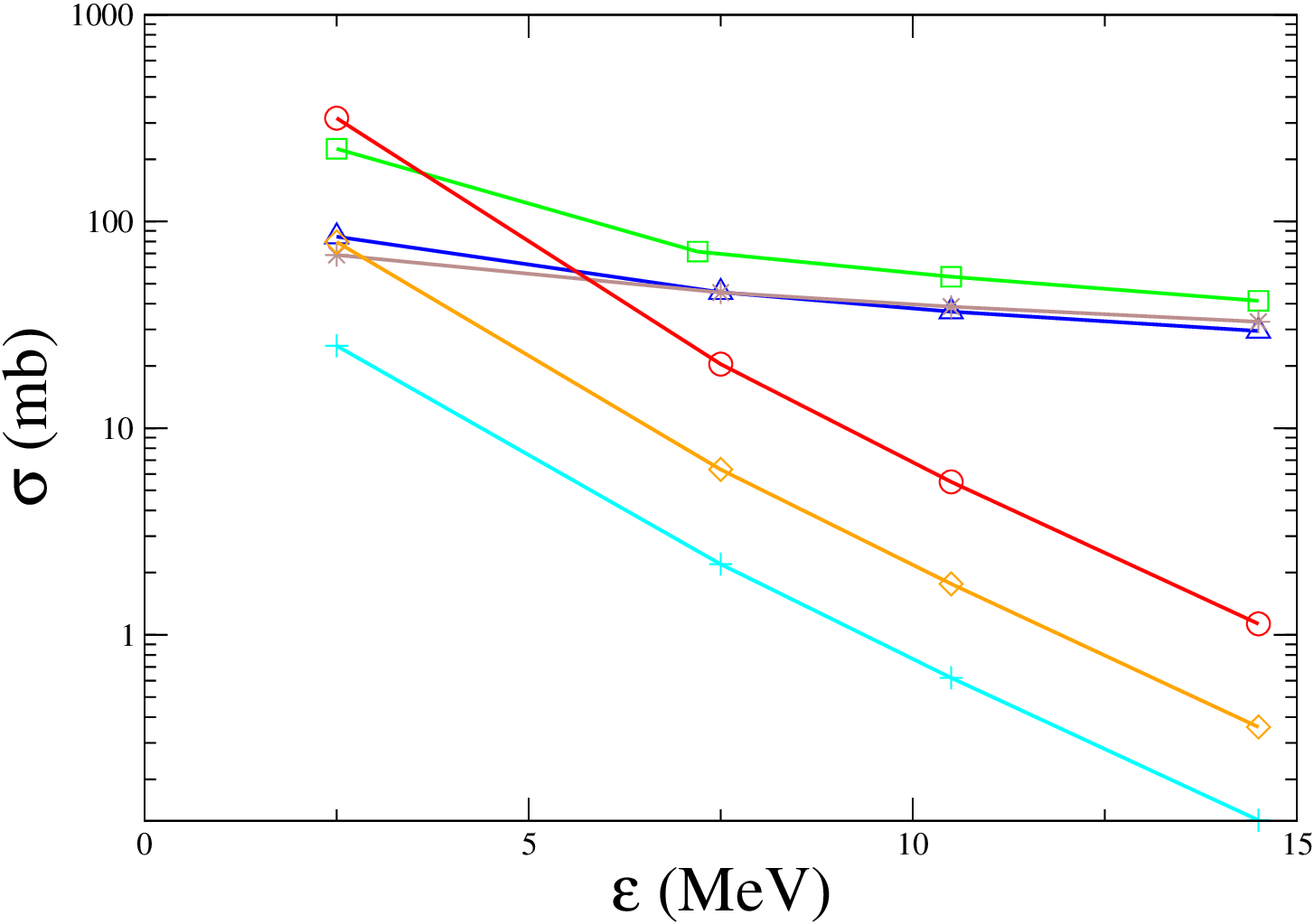}}
               \caption{ \footnotesize Integrated  breakup cross-section for a hypothetical $^{34}$Si beam against Pb at 70 A.MeV as a function of the neutron separation energy. Different initial parameters: circles (squares) are for Coulomb (nuclear) breakup from an initial s-wave; diamonds (triangles) for Coulomb (nuclear) breakup from a d-wave; pluses (stars) for Coulomb (nuclear) breakup from an initial f-wave. Nuclear breakup is the sum of diffractive and stripping contributions}
\label{fig3}
\end{minipage}
\hspace{\fill}
\begin{minipage}[t]{75mm}
\scalebox{0.27}{
               \includegraphics{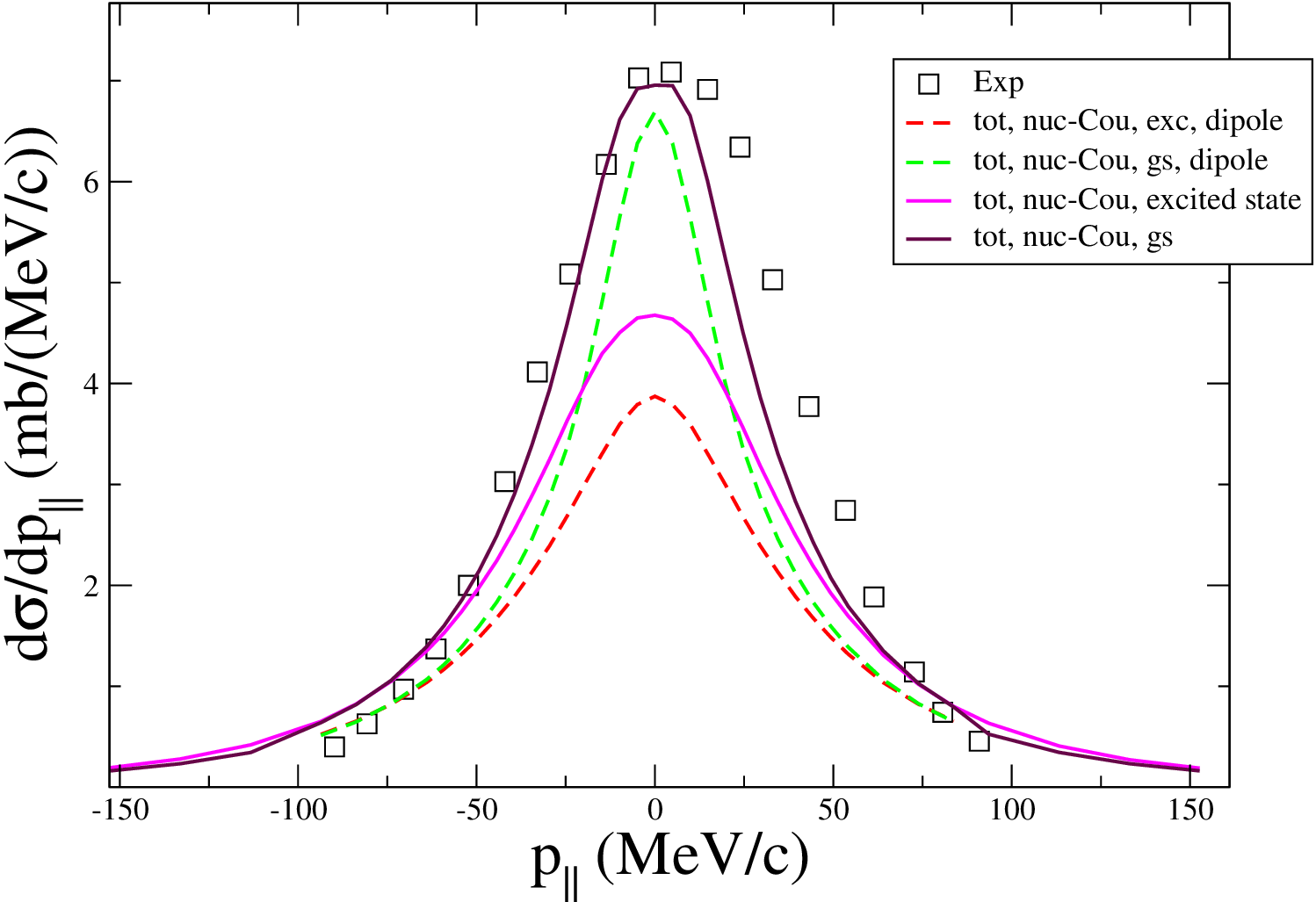}}
\caption{\footnotesize Calculated  momentum distribution of $^7$Be fragments after proton-removal from $^8$B against Pb at 936 MeV/A. Both dipole and full multipole results are shown for the ground state and first excited state.  Calculations according to Sec.3 and Ref.\cite{noip} where more details can be found. Data are from Ref.\cite{lola03}.}
\label{fig4}
\end{minipage}
\end{figure}

\subsection{Sudden limit and all-order treatment}

 Aiming for an all-order formalism, in \cite{1} it was shown that a possible way to achieve this is to use the sudden approximation, subtract the first order term, which diverges for large impact parameter, and then to add a first order term calculated in time-dependent perturbation theory.
The sudden limit ($\omega \to 0$) must be therefore taken in the above expression for $\chi_{eff}$, yielding $\chi_{eff}^{sudd}=\frac{2 V_C}{\hbar v} \log{\frac{b_c}{R_\perp}}.$
 Following a procedure analogous to that of \cite{nois}, the Coulomb phase for the proton is shown to be \cite{noip}
\begin{eqnarray} \label{fasipr}
\chi^p=\frac{2}{\hbar v}\left(V_c e^{i\beta_1 \omega z /v}K_0(\omega b_c/v) -V_C K_0(\omega R_\perp/v)+V_v e^{-i\beta_2 \omega z /v}K_0(\omega b_v/v)\right)
\end{eqnarray}
 Since $V_C=V_c+V_v$, eq. (\ref{fasipr}) can be written as
$
\chi^p=\chi(\beta_1,V_c)+\chi(-\beta_2,V_v)
$
 where $\chi$ is the $\chi_{eff}$ of the previous section and ${ \bf b}_v={\bf b}_c+{\bf r}_{\perp}$ is the proton impact parameter with respect to the target. The Coulomb phase is therefore the sum of two terms: one of them describes the recoil of the core whereas the other accounts for the direct proton-target Coulomb interaction. Of course, in the case of the neutron the latter vanishes and the phase reduces to the one derived in \cite{nois}. 
 It is easy to see that the expansion of $\chi_{eff}$ to first order in ${\bf r}$ yields the well known dipole approximation to the phase:
 which only differs from the neutron breakup case in the different constant factor, which is now $(V_c\beta_1-V_v\beta_2)$ instead of $V_c\beta_1$ of Ref. \cite{nois}.

 The probability amplitude can be written as the sum of three contributions. The recoil term,
\begin{equation}
A^{rec}=\int d{\bf r} e^{-i {\bf k} \cdot {\bf r}} \phi_i({\bf r}) \left( e^{i \frac{2 V_c}{\hbar v}\log{\frac{b_c}{R_\perp}}} -1 -i\frac{2 V_c}{\hbar v}\log{\frac{b_c}{R_\perp}} +i\chi(\beta_1,V_c) \right),
\label{rec}\end{equation}
 where, according to the discussions in \cite{1,nois}, the sudden limit has been used in order to include all orders in the interaction. The direct proton Coulomb interaction term A$^{dir}$ which has the same form as Eq.(\ref{rec}) but with the substitution V$_c\to$V$_v$, $\beta_1\to-\beta_2$ and b$_c\to $b$_v$. 
The nuclear part is given by
\begin{equation} \label{nuq}
A^{nuc}=\int d{\bf r} e^{-i{\bf k}\cdot {\bf r}} \left( e^{i \chi_{nt}(b_v)}-1\right) \phi_{i}({\bf r}),
\label{nuclA}\end{equation}
which is the well known eikonal form of the diffractive nuclear breakup
with the neutron-target phase $
\chi_{nt}({\bf b}_v)=\int dz  V_{nt}({\bf b}_{v},z) /(v\hbar)
$.
Finally the expression for the differential cross-section is
\begin{eqnarray}
\frac{d \sigma}{d {\bf k}}=\frac{1}{8\pi^3}\frac{m k}{\hbar^2}\int d {\bf b}_c |S_{ct}(b_c)|^2 |A^{nuc}+A^{dir}+A^{rec}|^2.
\end{eqnarray}
In a number of papers higher order effects \cite{esben95} and proton breakup have been discussed, among which we recall Refs.\cite{kelly,hen96,ang04}. Some works have also addressed the problem of asymmetry in the core parallel momentum distribution after proton knockout \cite{esben96,davids01}. The fact that this asymmetry comes from high order terms can be directly extracted from our formalism. If the Coulomb part of the amplitude is simply expanded to first order in $\chi$, it can be written, in terms of the one-dimensional Fourier transform in $z-$direction $\hat{\phi}_i$, as
\begin{eqnarray}
A^{Cou}&\simeq &\int d{\bf r}_\perp e^{-i {\bf k}_\perp \cdot {\bf r}_\perp} \frac{2}{\hbar v}\left(V_c K_0(\omega b_c/v)\hat{\phi}_i({\bf r}_\perp,k_z-\beta_1 \omega/v) -V_C K_0(\omega R_\perp/v)\hat{\phi}_i({\bf r}_\perp,k_z) \right. \nonumber \\
 &+&\left.V_v K_0(\omega b_v/v)\hat{\phi}_i({\bf r}_\perp,k_z+\beta_2 \omega/v) \right).\label{asy}
\end{eqnarray}
 Thus the Coulomb breakup probability amplitude can be regarded as a coherent sum of three terms, each of which contains a shifted $z-$Fourier transform. The shifts are in opposite directions, $\beta_1 \omega/v$ and $-\beta_2 \omega/v$, but they are not visible directly in the calculated momentum distributions as $\omega$ depends on $k$ itself. Moreover, the $1/v$ factor indicates that the asymmetry decreases as the beam energy increases.
In the dipole  approximation, however, the amplitude  does not contain any asymmetry for the momentum distribution as it involves square modulii of $\hat{\phi}_i({\bf r}_\perp,k_z)$ separately. Hence we have confirmed analytically that the asymmetry in Coulomb breakup parallel momentum distributions is due to the presence of higher multipole terms, in agreement with earlier works \cite{esben95,esben96,davids01}. However, the presence of the nuclear interaction introduces an interference that does depend on the sign of $k_z$ and thus an additional asymmetry to that due to higher multipole terms in the Coulomb interaction.
 
We then present two applications of the formalism just discussed. Fig. 1  shows calculations of absolute cross sections for Coulomb and nuclear breakup, according to the formalism of Secs. 2 and 3, for a heavy exotic projectile $^{34}$Si. It intends to demonstrate the feasibility of nuclear breakup experiments on heavy targets when the initial neutron separation energy and angular momentum become large. In this way it should be possible to avoid the asymmetries and deviations from the eikonal model found in some experiments \cite{enders01,gade05}. Fig. 2 compares data \cite{lola03} for proton breakup to calculations from our new model \cite{noip} for proton breakup shortly described above.
 
\section{Transfer to the Continuum}
 In Ref.\cite{bb} the transfer to the continuum method (TC) to calculate the nuclear breakup was introduced in a way that  made  numerical calculations relatively easy.   Furthermore it was shown that breakup gives rise  to  a stripping cross section $\sigma_{str}$ and  a
diffractive breakup cross section $\sigma_{diff}$ which are distinguishable experimentally depending on whether
the removed neutron is detected in the final state or not.    
The eikonal approximation to Eq.(\ref{1}) has been already given for the diffraction term by Eq.(\ref{nuclA}). Extending it to the stripping term of the nuclear breakup \cite{6,bon01},  one  derives \cite{bon98a} the total one nucleon removal probability
\begin{equation}{dP_{-n}(b_c)\over d k_z}\sim {1\over 2\pi}
\int_0^\infty d{\bf b_v}  \left[|(1-e^{-i\chi({ b_v})})|^2 +
 1-|e^{-i\chi({
b_v})}|^2\right]
 |\tilde {\phi}_i ({\bf  b_v-b_c}, k_z)|^2 ,\label{pg}\end{equation}
where    $ e^{-i\chi({ b_v})}$ is
the eikonal form of the neutron  (proton) target S-matrix already discussed in Sec.3. Notice that in this expression the exact initial state wave function appears, therefore Eq.(\ref{pg}) is valid for a neutron as well as for a proton.
 $|\tilde {\phi}_i({\bf b_v-b_c}, k_z)|^2$ is  the
longitudinal Fourier transform of the initial state wave function. The total breakup probability is obtained from the integral of
Eq.(\ref{pg}) involving
$
{\cal I}(k_z^{min},k_z^{max}) =\int_{k_z^{min}}^{k_z^{max}}d k_z \vert\bar \phi_i({\bf
b_v-b_c},k_z)\vert^2.
$
If the integral could be extended to $\pm\infty$, it would just be the longitudinal
density, and the formulae for the TC and eikonal model would become
identical. In fact in this limit
the removal cross section  reduces to
\begin{equation}\sigma_{-n}=C^2S
\int d^2{\bf b_c}\int d^3{\bf r}\left[|(1-\bar S)|^2 + 1-|\bar
S|^2\right] |S_{ct}(b_c)|^2|{\psi}_i ( {\bf
b_v-b_c},z)|^2\label{creik}\end{equation}
 which is consistent with the 
breakup cross section originally obtained by Yabana and collaborators  \cite{yab}. Notice that  Eqs.(\ref{pg}) and (\ref{creik}) are consistent with eq.(\ref{nuclA}).
 
  To see
how accurate the sudden approximation is, the
integral ${\cal I}(k_z^{min},k_z^{max})$ was calculated in Ref.\cite{bon01} under various conditions of angular momentum, neutron
binding energy in the projectile, and projectile velocity.  For values of the parameters
of interest there can be a rather large reduction for small values of the
neutron transverse radius in the projectile, $|b_v-b_c|$.  However,
the approximation becomes increasingly accurate as the transverse
radius is made larger. Another sudden model for Coulomb and nuclear breakup was presented in Ref.\cite{4} and compared to the present approach.

\section{Projectile Fragmentation}
We call {\it projectile fragmentation}  the  elastic breakup (diffraction dissociation) discussed above, when the observable studied is the neutron-core relative energy spectrum. This kind of observable has been widely  measured in relation to the Coulomb breakup on heavy target. Results on light targets have also been presented  \cite{fuku}. These data enlighten the effect of the neutron final state interaction with the core of origin, neglected in the previous sections, while observables like the core energy or momentum distributions enlighten the effect of the neutron final state interaction with the target.

Projectile fragmentation has also been used experimentally  with
  two neutron halo projectiles \cite{nig}-\cite{Labi99}. In this case  it has been suggested that the  reaction might proceed by  
the simultaneous emission of the two neutrons or by
successive emissions \cite{nig}. The successive emission can be due to a mechanism in which one neutron is stripped by  the interaction with the target, as in the one-neutron fragmentation case, while the other is left behind, for example in a resonance state, which then decays. This mechanism has been described by the sudden approximation  in Ref.\cite{bhe} under the hypothesis that while the first neutron is stripped, the second neutron is emitted at large impact parameters with no final state interaction with the target. The emission can be expected sequential if the two neutrons are not strongly correlated.

If the two neutrons are strongly correlated they will preferentially be emitted simultaneously.  If  the neutron which is not detected is stripped while the other suffers an elastic scattering on the target, then
in both cases to first order in the interaction the neutron  ends-up in a plane wave final state \cite{bb}. It can then re-interact with the core which, for example,  is going to be $^{10}$Be in the case of the one-neutron halo projectile $^{11}$Be, while it will be  $^{12}$Be in the case of the projectile fragmentation of $^{14}$Be, since $^{13}$Be is not bound. 
Experiments with a  $^{14}$B projectile \cite{jl} have also been performed, in which the n-$^{12}$Be relative energy spectra have been reconstructed by coincidence  measurements. In such a nucleus the valence neutron is weakly bound, while the valence proton is strongly bound. Thus the neutron will probably be emitted in the first step and then re-scattered by the core minus one proton nucleus. The projectile-target distances at which this kind of mechanism would be relevant are probably not so large to neglect the effect of the neutron-target interaction.  
 \subsection {Inelastic excitation to the continuum. }

To first order the inelastic-like excitations can be described again by the  
time dependent
perturbation amplitude Eq.(\ref{1}) \cite{adl,bb}.
 In this section also, the potential $V({\bf { r,R}}(t))$, which  is the interaction responsible for the neutron transition,
 moves past on a constant velocity path as described in the previous sections.     The  radial part $ \phi_{i}({\bf r})$  of  the single particle initial state wave function $\psi_{i} ({\bf r},t)$ is calculated in a potential $V_{WS}( r)$  which is  
fixed in space.
The coordinate system  and other details  of the calculations can be found in  Ref.\cite{5}.  
 In the special case of exotic nuclei
the traditional approach to inelastic excitations needs to be modified.    For example the final state can be eigenstate of a potential $V_1$
modified with respect to $V_{WS}$  because some other particle is  
emitted during the reaction process as discussed in the introduction. The final state interaction might also have  
an imaginary part which would take into account the  
coupling between
a continuum state and an excited core. 
The first order time dependent perturbation amplitude then reads
\begin{equation}
A_{fi}={1\over i\hbar v}
\int_{-\infty}^{\infty}dx dy dz ~ \phi^*_{f}
(x,y,z)\phi_{i}(x,y,z){e^{iqz}}\tilde V(x-b_c,y,q),\label{c}
\end{equation}
where
$\tilde V(x-b_c,y,q)= \int_{-\infty}^{\infty}dz V(x-b_c,y,z)e^{iqz},$ and we changed  variables and put $z^{\prime}=z - vt$ or $t = (z -  
z^{\prime})/v$,
$
q={{\varepsilon_f-\varepsilon_i}/ {\hbar v}}
$. In this section $\varepsilon_f$ is the neutron-core relative energy in the final state.

The target represented by $\tilde V$  perturbs the initial bound state wave function and allows the transition to the continuum by transferring some momentum to the neutron. Then it is enough to choose a simplified form of the interaction,  such as a delta-function potential
$V(r) = v_2\delta(x)\delta(y)\delta(z)$. The value of the strength  $v_2\equiv$ [MeV fm$^3$]
 used in the calculation is taken equal to the volume integral of the appropriate neutron-target interaction. It is clear that  while in the sudden approach the initial and final state overlap is taken in the whole coordinate space, irrespective of the target and of the beam velocity, here 
the overlap of the initial and final wave functions depends on the core-target impact parameter. The neutron is emitted preferentially on the reaction plane  and the z-component, being along the relative velocity axis is boosted by a momentum $q$.

\begin{figure}[htb]
\begin{minipage}[t]{80mm}
\scalebox{0.28}{
               \includegraphics{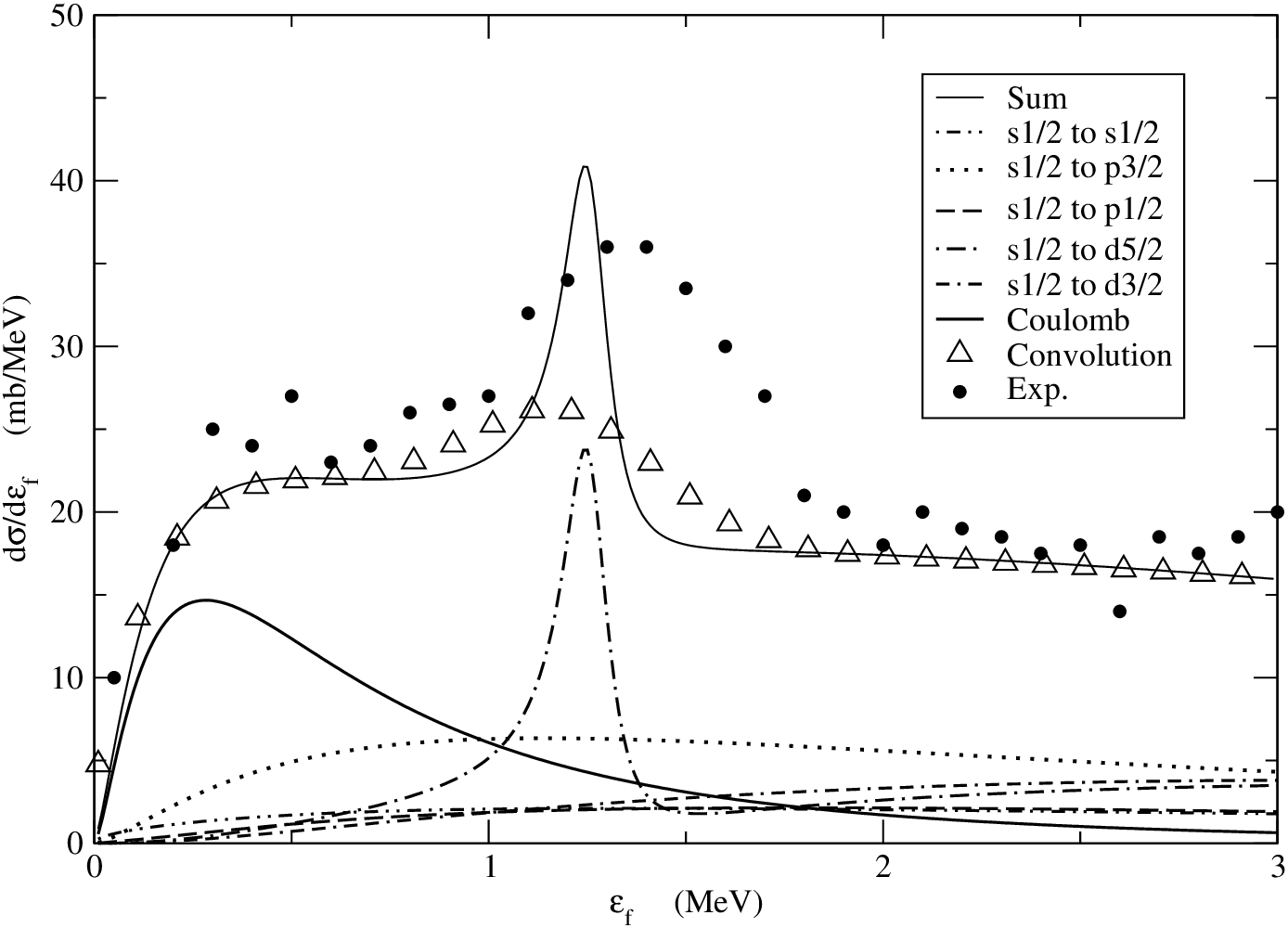}}
\caption{ \footnotesize n-$^{10}$Be relative energy  spectrum,
including Coulomb and nuclear breakup for the reaction  $^{11}$Be+$^{12}$C $\to$ n+$^{10}$Be+X at 69 A.MeV. Only the contributions from an s initial state with  
spectroscopic factor C$^2$S= 0.84 are  calculated. The  triangles are the total calculated result after convolution with the experimental resolution function. The dots are the experimental points from \cite{fuku}. }
\label{fig1}
\end{minipage}
\hspace{\fill}
\begin{minipage}[t]{75mm}
\scalebox{0.28}{
               \includegraphics{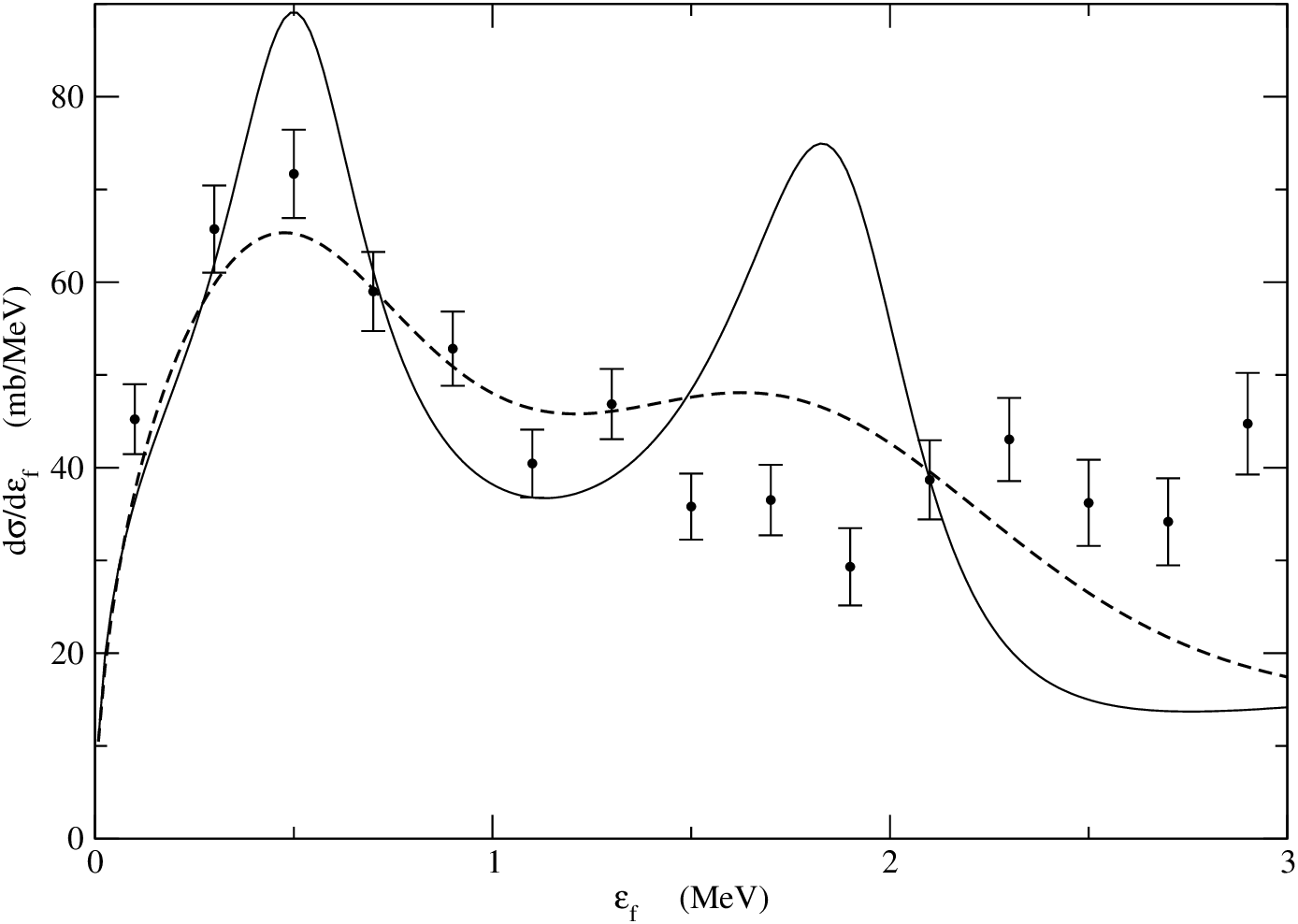}}
\caption{ \footnotesize Sum  
of all transitions from the s initial state with $\varepsilon_i$=-1.85 MeV (solid line)   for the reaction $^{14}$Be+$^{12}$C $\to$ n+$^{12}$Be+X . Experimental points from H. Simon et al.  \cite{simo} for the same reaction at 250 A.MeV. Dashed line is the folding of the calculated spectrum with the experimental resolution curve.}
\label{fig2}
\end{minipage}
\end{figure}
Due of the strong core absorption discussed in Secs. 2 and 3 
these calculations are also performed using the asymptotic form of the initial and final state wave functions. 
Introducing the quantization condition
 according to Ref.\cite{bb}
the probability spectrum reads
\begin{eqnarray}
{dP_{in}\over d\varepsilon_f}={2\over \pi}{v_2^2\over \hbar^2  
v^2}{C_i^2 }{m\over\hbar^2k}{1\over 2l_i+1}\Sigma_{m_i,m_f}
 |1-\bar S_{m_i,m_f}|^2 |I_{m_i,m_f}|^2. \label{8}
\end{eqnarray}
 The generalization including spin is given in Appendix B of Ref.\cite{5} and   $  |I_{m_i,m_f}|^2=\left|\int_{-\infty}^{\infty} dze^{iqz}i^{l_i} \gamma h^{(1)}_{l_i}(i\gamma  
r)Y_{{l_i},{m_i}}(\theta,0)  k{i\over 2}h^{(-)}_{l_f}(kr)Y_{{l_f},{m_f}}(\theta,0)\right|^2.$
The quantity $\bar S=e^{2i(\delta+\nu)}$ is an off-the-energy-shell S-matrix representing the final state interaction  
of the neutron with the projectile core.  It depends on a phase which is the sum of $\delta$, the free particle n-core phase shift, plus  $\nu$ the phase of the matrix element  $|I|$.  Two examples of our calculations \cite{5}
are shown in Figs. 3 and 4  and compared to recent data.  See also Ref.\cite{3}.
Finally we refer to two books \cite{01,02} and recent reviews \cite{03,04} in which further discussion and bibliography can be found.

\section{Conclusions and Outlook}
The field of Rare Isotopes Studies is very active, growing steadily  and rapidly. Some recent achievements in reaction theories for breakup and elastic scattering of exotic beams have been presented.   From the structure point of view, in the search for the dripline position, a very important role is played by the study of nuclei unstable by neutron emission. On the other hand increasing the mass of the projectiles produced we are going to face the problem of envisaging new experiments to study them. These two are among the most important subjects which need to be adressed and further developed in the near future and for which some suggestions have been presented.


\begin{thebibliography}{99}
 \bibitem{adl} K. Alder and A. Winther,  Electromagnetic Excitation, North-Holland, 1975.
\bibitem{bb} A. Bonaccorso and D.M. Brink, Phys. Rev. C 38 (1988) 1776; Phys. Rev. C 43 (1991) 299;
 Phys. Rev. C 44 (1991)  1559.

 \bibitem{bb4} A. Bonaccorso, Phys. Rev. C 60 (1999) 054604.
 \bibitem{1} J. Margueron, A. Bonaccorso and D.M. Brink, \np{A 703} (2002) 105;  Nucl. Phys.  A 720 (2003) 337.
\bibitem{bw} R. A. Broglia and A. Winther,  
Heavy Ion Reactions, Benjamin, Reading, Mass, 1981.

\bibitem{kox} S. Kox et al., Phys. Rev. C 35 (1987) 1678.
\bibitem{nois} A. Garc\'ia-Camacho, A. Bonaccorso and D.M. Brink,
Nucl. Phys.   A 776 (2006) 118. 
\bibitem{noip} A. Garc\'ia-Camacho, G. Blanchon,  A. Bonaccorso and D.M. Brink, in preparation.

\bibitem{esben95} H. Esbensen, G.F. Bertsch and C. Bertulani, \np{A 581} (1995) 107.

\bibitem{kelly}   J.H. Kelley et al., Phys. Rev. Lett. { 77} (1996) 5020 and references therein.

\bibitem{hen96}   K. Hencken, G. Bertsch and H. Esbensen, \pr{C 54} (1996) 3043.

\bibitem{ang04} A. Bonaccorso, D. M. Brink and C. A. Bertulani, \pr{C 69} (2004) 024615..


\bibitem{esben96} H. Esbensen and G.F. Bertsch, \np{A 600} (1996) 37; Phys.  Rev. C { 59}  (1999) 3240; \np{A 706} (2002) 383.



\bibitem{davids01} B. Davids et al., Phys. Rev. Lett. {81} (1998) 2209;  \pr{C 63} (2001) 0654806.


\bibitem{enders01} J. Enders et al., \pr{C 65} (2001) 034318.

\bibitem{gade05} A. Gade et al., \pr{C 71} (2005) 051301.

  \bibitem{lola03} D. Cortina-Gil et al., \np{A 720} (2003) 3.

\bibitem{6} A. Bonaccorso and F. Carstoiu, Phys. Rev. C 61  (2000) 034605.
\bibitem{bon01} A. Bonaccorso and G.F. Bertsch, \pr{C 63} (2001) 044604.
\bibitem{bon98a} A. Bonaccorso and D.M. Brink, \pr{C 57} (1998) R22; \pr{C 58} (1998) 2864.

\bibitem{yab} K. Yabana, Y. Ogawa and Y. Suzuki, Nucl. Phys. { A 539}  (1992) 295.


\bibitem{4} F. Carstoiu, E. Sauvan, N.A. Orr, A. Bonaccorso.
 Phys. Rev.  C 70 (2004) 054602.

\bibitem{fuku} N. Fukuda et al., Phys. Rev. C 70,  (2004) 054606.
 






\bibitem{nig}
F.M. Marqu\' es et al., Phys. Rev. C 64 (2001) 061301(R). 
N.Orr, Prog. Theor. Phys. Suppl. 146  
(2003) 201.
\bibitem{bhe} G.F. Bertsch, K. Hencken and H. Esbensen, Phys. Rev. C 57  (1998) 1366.
\bibitem{jl} J.L. Lecouey, Few Body Syst. 34 (2004) 21-26. 

\bibitem{5} G. Blanchon, A. Bonaccorso, D.M. Brink, A.Garc\'ia-Camacho and N. Vinh Mau.
{\it Unbound exotic nuclei  studied by projectile fragmentation}, submitted to NPA.

 \bibitem{3} G. Blanchon, A. Bonaccorso and N. Vinh Mau,
 Nucl. Phys.   A 739  (2004) 259.
\bibitem{mt} M. Thoennessen et al., Phys. Rev. C 59 (1999) 111; Phys. Rev. C 60 (1999) 027303; Proceedings the Erice Int. School of  
Heavy-Ion Physics, 4$^{th}$ Course. Eds. R.A. Broglia and P.G. Hansen. World Scientific,  Singapore 1998, pag.269.

\bibitem{labi} M. Labiche et al.  Phys. Rev. Lett.  86  (2001) 600.


\bibitem{simo} H. Simon et al., Nucl. Phys. A734 (2004) 323 and private communication.


\bibitem{Kor95} A.A. Korsheninnikov et al., Phys. Lett. B 343 (1995)  
53.



\bibitem{Bert91} G.F. Bertsch and H. Esbensen, Ann. Phys. (N.Y.) 209  
(1991) 327.
\bibitem{Thomp96} I.J. Thompson and M.V. Zhukov, Phys. Rev. C 53  
(1996) 708.
\bibitem{Vinh96} N. Vinh Mau and J.C. Pacheco, Nucl. Phys. A 607 (1996)  163.
\bibitem{pa} J.C. Pacheco and N. Vinh Mau, Phys. Rev. C 65  
(2002) 044004.
\bibitem{de} P. Descouvemont, Phys. Lett. B 331 (1994) 271; Phys. Rev. C 52 (1995) 704.

\bibitem{ba} A. Adachour, D. Baye and P. Descouvemont, Phys. Lett. B 356  (1995) 445.

\bibitem{bab} D. Baye, Nucl. Phys. A 627 (1997) 305.
\bibitem{ta} T. Tarutina, I.J. Thompson, J.A. Tostevin, Nucl. Phys.  
A 733 (2004) 53.



\bibitem{Labi99} M. Labiche, F.M. Marques, O. Sorlin and N. Vinh Mau,  Phys. Rev. C 60 (1999) 027303.



\bibitem{01}
   C.A. Bertulani, M.S. Hussein, G. M\"unzenberg. Physics of Radioactive Beams. Nova Science Publ. 2001.
     \bibitem{02} Y. Suzuki,  R. G. Lovas, K. Yabana,  K. Varga.
Structure  and Reactions of Light  Exotic  Nuclei.  Taylor and Francis Eds. 2003.
\bibitem{03} L.F. Canto, P.R.S. Gomes, R. Donangelo, M.S. Hussein, Phys. Rep. 424 (2006) 1.

\bibitem {04} B. Jonson, Phys. Rep. 389 (2004) 1.



 

\end{thebibliography}
\end{document}